
%
\input harvmac

\def \rd {\partial}

\def \a {\alpha}
\def \b {\beta}
\def \d {\delta}
\def \D {\Delta}
\def \g {\gamma}
\def \ld {\lambda}

\def \O {\Omega}
\def \o {\omega}
\def \P  {\Psi}
\def \t {\tau}
\def \th {\theta}
\def \Th {\Theta}
\def \v {\vert}

\def  \ra {\rightarrow}

\def \h {\hbar}
\def \bd#1 {{\bf #1}}
\def \ke {\rangle}

\Title{KYUSHU-HET-3}{ Berry's Phase and Euclidean Path Integral}
\centerline {Taro KASHIWA, Shuji NIMA and Seiji SAKODA}
\bigskip
\centerline {Department of Physics}
\centerline{Kyushu University}
\centerline {Fukuoka 812, Japan}
\vskip 1cm
A method for finding Berry's phase is proposed under
the Euclidean path integral formalism. It is characterized by picking up
the imaginary part from the resultant exponent. Discussion is made
on the generalized harmonic oscillator which is shown being so universal
in a single degree case. The spin model
expressed by creation and annihilation
operators is also discussed. A systematic way of expansion in the
adiabatic approximation is presented in every example.

\Date{1/92}
\newsec{Introduction}
There have been intensive studies
on Berry's phase\ref\Bea{M. V.Berry, Proc. R. Lond.
{\bf A392} (1984), 45.}\ref\Book{A. Shapere and F. Wilczek (ed.), ``Geometric
Phases in Physics," World Scientific, Singapore, 1989.}  as well as other
geometric phase\ref\AA{Y. Aharonov and J. Anandan, Phys. Rev. Lett. {\bf 58}
(1987), 1593.}. However as far as the path integral formalism is concerned,
there still lacks a general as well as a systematic consideration. A typical
example can be seen as follows:
let $|n({\bf  X} (t)) \rangle$ be eigenstate, with
eigenvalue $E_n({\bf X}(t))$, of some time-dependent Hamiltonian  $ H({\bf X}
(t))$:       \eqn\ada{
H(\bd{X} (t))\v n(\bd{X} (t)) \ke=E_{n}(\bd{X} (t))\v n(\bd{X} (t))\rangle
\quad,   }
\eqn\adaa{
\langle n(\bd X (t))\v m(\bd X (t))\rangle=\d_{n,m} \quad,
}
\eqn\adaaa{
\sum_{n}\v n(\bd X (t))\rangle\langle n(\bd X (t))\v=\bd 1 \quad.
}

The Schr\"odinger equation,
\eqn\adb{
i\h{\rd \over \rd t} \v \P (t)\ke=H(\bd{X} (t))\v \P (t) \ke\quad .
}
brings the time evolution operator, $U(t)$, defined by
\eqn\adbb{
 |\Psi(t)\ke = U(t)|\Psi(0)\ke\quad ,
}
to
\eqn\adba{
i\h{\rd \over \rd t}U(t) = H(\bd{X} (t))U(t),\quad U(0) = \bf{1}\quad .
}
The solution of $\adba$ is given by
\eqn\add{
\eqalign{
U(T) &= \lim_{N \ra \infty
}\left(1-{i \over \h}\D tH(\bd {X}
(t_{N-1}))\right)\left(1-{i \over \h}\D tH(\bd {X} (t_{N-2}))\right) \cr
\ \ \ \ \ \ \ \ \ &\times\cdots\times\left(1-{i \over \h}\D tH(\bd {X}
(t_{1}))\right)\left(1-{i \over \h}\D tH(\bd {X} (t_{0}))\right) \quad \cr
&\equiv \bd {T} \exp \left( -{i\over \h} \int_{0}^{T} dt H(\bd {X} (t))
\right) \quad , \cr}
}
where
$\Delta t={T/ N},\ t_k=k\Delta t$ $(k=0,\cdots, N)$, with $ t_0=0, t_N=T .$
If we would insert $\adaaa$ into $\add$ successively and use
\eqn\wrong{
\langle n({\bf X}(t))|m({\bf X}(t-\Delta t))\rangle \approx{\delta
}_{n, m} {\left(1-\Delta t\langle n({\bf X}(t))|{\partial  \over \partial
t}| n({\bf X}(t))\rangle \right)}\quad,
}
we could get
$$
U(T) \approx \exp {\Bigl[ \int_{0}^{T}dt \left( -{i\over\hbar} E_n({\bf X}
(t) )- \langle n({\bf X}(t))|{\partial  \over \partial t}| n({\bf X}(t))\rangle
\right) \Bigr]} \quad,
$$
and identify the first(second) term as the dynamical (geometrical, see
below) part in the exponential\ref\Kuratsuji{ H. Kuratsuji and S. Iida, Prog.
Theor. Phys. {\bf 74} (1985), 439.  M. Stone, Nucl. Phys.  {\bf B327}
(1989), 399.}. It is not allowed, however, to put $\wrong$ except in the
semi-classical approach. Thus as for the geometrical object it
may be legitimate
but there is no way for higher order corrections if we assume $\wrong$.

 Instead the adiabatic approximation should be adopted\ref\Mes{A. Messiah,
``Quantum Mechanics," Vol.II, p.739, North-Holland, Amsterdam, 1970.} under
the assumption  that there will be no level crossing:  first we make a
time-dependent unitary transformation, such that
\eqn\ade{
V^\dagger (t)H(t)V(t) = H^V(t)\quad ,
}
where a diagonal $H^V(t)$ is given by
\eqn\adf{
H^V(t) =\sum_n E_n(t)|n(\bd {X} (0))\rangle\langle n(\bd {X} (0))|\quad .
}
Here and in the following we write $H(t)(E_n(t))$ instead of $H(\bd{X} (t))
 (E_n(\bd {X} (t) ))$. Substituting $\ade$
(in terms of $H=VH^VV^\dagger$) into
\ada\ and using \adf, we have
\eqn\adff{
V^\dagger(t)|n(\bd {X} (t))\ke = e^{-i\gamma_n(t)}|n(\bd {X} (0))\ke\quad,
}
where $\g_n$ is some phase. Since there is an arbitrariness
in $\g_n$'s definition, we
impose the `off-diagonal condition';
\eqn\adg{
\langle n(\bd {X} (0))|V^\dagger(t)\dot V (t)|n(\bd {X} (0))\rangle = 0\quad,
}
which leads us to
\eqn\adh{
\eqalign{
\g_n(t)&=i\int_{0}^{t}dt'\langle n(\bd X (t'))\v%
{\rd \over \rd t'}\v n(\bd X (t'))\rangle
\cr
     &=i\int_{\bd X (0)}^{\bd X (t)}dX^{a}%
\langle n(\bd X )\v{\rd \over \rd X^{a}}
\v n(\bd X )\rangle\quad. \cr
} }
This phase, $\g_n$, is called as a geometric phase and for
some $T$, such that ${\bf
X}(T)={\bf X}(0)$ and $|n({\bf X}(T)\rangle=|n({\bf X}(0)\rangle$,
it is nothing but
Berry's phase \Bea\ \Book. Inserting $\adaaa$ (with putting $t=0$) into \add\
successively then using \ade, we find
\eqn\au{
\eqalign{
U(T) &= \lim_{ N\rightarrow \infty}\sum_{
\scriptstyle \lbrace n_k\rbrace  \atop
k=0,\cdots, N } V(t_N)|n_N(0)\rangle\langle n_N(0)| \cr
&\times\left(1-{i\over
\hbar}\Delta t \left\{H^V(t_{N-1})-i\hbar V^\dagger(t_{N-1})\dot
V(t_{N-1})\right\}\right) |n_{N-1}(0)\rangle\langle n_{N-1}(0)| \cr
&\times\cdots\times\left(1-{i\over
\hbar}\Delta t \left\{H^V(t_0)-i\hbar V^\dagger(t_0)\dot
V(t_0)\right\}\right)|n_0(0)\rangle\langle n_0(0)|\quad , \cr}
}
which is further rewritten, by use of the `off diagonal condition' \adg, as
\eqn\adi {
\eqalign{
U(T) = &\lim_{N\rightarrow\infty}\sum_{\lbrace n_k \rbrace}
V(t_N)|n_N(0)\rangle\langle n_N(0)|\left(1-{
i\over\hbar}\D t H^V(t_{N-1})\right)
|n_{N-1}(0)\rangle\cr
&\times \langle n_{N-1}(0)| \times\cdots\times\left(1-{i\over\hbar}\Delta
t H^V(t_0)\right)|n_0(0)\rangle\langle n_0(0)| \Bigl\{1+O(1/T)\Bigr\} \cr
= & \sum_n \exp \left(-{i\over\hbar} \int_{0}^{T}dt E_n(t)+i\gamma_n(T)
\right)|n(T)\rangle\langle n(0)| \Bigl\{1+O(1/T)\Bigr\}, \cr}
}
where we have used \adf\ and \adff\ to get the final result. Note that
contrary to $\wrong$ the state remains the same energy level all the time
because of the adiabatic approximation: if there is a change, from $|n({\bf
X}(0))\ke$ to $|m({\bf X}(0))\ke$ at $t={t}_m$, the contribution
is negligible;
$$
\eqalign{
     \int_{0}^{T} d{t}_{m}  & \langle m(0)| V^\dagger(t_m)
                                     \dot V(t_m) | n(0) \rangle     \cr
   & \times {\rm exp} \bigl[ {-{i \over
\hbar} \int_{{t}_{m}}^{T} dt ({E}_{m}(t)-{E}_{n}(t))} \bigr]
    \approx O({1\over T} )  \quad. \cr
}
$$

Hence the Feynman kernel, $K(q,q';T) \equiv \langle q|U(T)|q'\rangle $,
is given by
\eqn\adk {
\eqalign{
K(q,q';T) =& \sum_n \psi_n(q;\bd {X} (T)) \psi_{n}^*(q';\bd {X} (0)) \exp
\left(-{i\over\hbar} \int_{0}^{T}dt E_n(t)+i\gamma_n(T) \right) \cr
&\times\Bigl\{ 1+O(1/T) \Bigr\} \quad , \cr}
}
where $\psi_n(q;\bd {X} (t)) (= \langle q|n(\bd {X} (t))\rangle)$'s are
instantaneous eigenfunctions of $H(\bd {X} (t))$.
It should be noticed that for
Berry's phase it is
necessary to find out the base  $ | n({\bf X}(0))\rangle$ and
actually it is rather hard except a few cases such as the two-level system.

While the left-hand-side of $\adk $ has a path integral expression (whose
derivation is  relegated to the appendix);
\eqn\dtk{
\eqalign{ K(q,q';T)
&=\lim_{N\ra\infty}\int \prod_{ i=1}^{N}{dp(i) \over {2\pi \h} }
\prod_{j=1}^{N-1} dq(j) \cr
&\times\exp \left( {i\over \h} \sum_{k=1}^{N} \Bigl \{ p(k)\D q(k)
  -\D tH(p(k),{\bar q}(k);{\bf X}(k)) \Bigr\} \right)
\Biggr\v_{q_{N}=q,q_{0}=q'}, \cr}
}
with $\Delta t\equiv T/N$
\eqn\df {
\eqalign{
\Delta q(k) &\equiv q(k)-q(k-1)\quad,\cr
{\bar q(k)} &\equiv {1\over 2}{(q(k)+q(k-1))}\quad.\cr}
}
By putting $\Delta s \equiv {\Delta t \over T}$ and $\Delta t \ra\ 0\ (N\ra
\infty)$, $\dtk$ becomes
\eqn\ctk{
K(q,q';T)=\int{\cal D}p{\cal D}q
          \exp\left( {iT\over \h}\int_{0}^{1}ds
          \left\{ {1 \over T}p {dq \over ds}
-H(p,q; \bd{X} (s)) \right\} \right)
\quad,
}
where we have employed a continuous representation
for notational simplicity.
{}From $\ctk$, we can see that the
adiabatic approximation, $1/T$-expansion with
$T\ra\infty$, is almost equivalent to the
WKB approximation; $\h$-expansion
with $\h\ra 0$. It is well-known that the
WKB  approximation is easily carried
out as the saddle point method under path integral.
Thus it might be convinced
that the path integral method is most suitable
to perform the  adiabatic
approximation. However it is still
difficult to pick up Berry's phase from the
resultant exponent; since in view of $\adk$ there
exists a dynamical part,
$\int_{0}^{T}dt E_{n}(t)$, as well as a
geometrical part in the exponent.

 In this paper, we wish to propose a general as well as a
systematic way for obtaining
Berry's phase under path integral. To pick up a geometrical
part from the exponent
exclusively, in the next section we discuss a `Euclidean
technique' in the path integral
formalism. The following two sections are devoted to check
our prescriptions.
The final section is the discussion and a rigorous treatment of the path
integral formula is relegated to the appendix.

\newsec{Basic Facts and the Euclidean Technique}

In this section, we propose a method of picking up Berry's phase from
the path integral formula. In order to
clarify our standpoint let us start with
a Hamiltonian
\eqn\genaa {
{\hat H}=H({\hat {\bf P}},{\hat {\bf Q}})+h({\hat
p},{\hat q};{\hat {\bf Q}}) \quad,
}
 where $({\bf {P,Q}})=(P_1,\cdots,P_a;Q_1,\cdots,Q_a); a\ge 2 $; whose masses
($M$'s generically) are much heavier than that of $(p,q)$'s. Here and in the
following the caret symbol is used to specify the quantum operator. $q$ moves
much faster than ${\bf Q}$'s
due to its lighter mass, thus is called as the fast
variable. ${\bf Q}$'s are then the slow variables. In terms of the Feynman
kernel this can be seen as follows:
\eqn\gena{
\eqalign{ K({\bf Q},q;&{\bf
Q}',q';t)\equiv\langle{\bf Q},q|
\exp\left(-{i\over \hbar}{\hat H}t\right)|{\bf
Q}',q'\rangle\cr &=\int{\cal D}{\bf P}{\cal D}{\bf Q}{\cal D}p{\cal
D}q\exp\left( {i\over \hbar}\int_0^t dt\left\{{\bf P}{\dot {\bf Q}}+p{\dot
q}-H({\bf {P,Q}})-h(p,q;{\bf Q})\right\}\right)\cr &\simeq K_{\scriptstyle
WKB,M\ra\infty}({\bf Q,Q'};t)K_{{\bf Q}c}(q,q';t)\quad,\cr}
}
where the third line is the cosequence of the heavy mass; since $M\ra\infty$
enables us to perform the WKB-approximation exactly with respect to $({\bf
{P,Q}})$ system. The resulting kernel is given by
\eqn\genc {
\eqalign{
 K_{{\bf Q}_{c}}&(q,q';T)  \cr
&\equiv\int{\cal D}p{\cal D}q\exp\Bigl[ {iT\over
\hbar}\int_0^1 ds \left\{{1\over T}p{\dot q}-h(p,q;{\bf
Q}_{c})\right\}\Bigr] \Bigg\v_{{\bf Q}_{c}(T)={\bf Q}_{c}(0)={\bf Q}_{0}},\cr
}}
where ${\bf Q}_{c}$ denotes the
classical solution of the $({\bf {P,Q}})$ system
with the boundary conditions
${\bf Q}_{c}(0)={\bf Q}', {\bf Q}_{c}(t)={\bf Q}$.
The effective Hamiltonian, $h({\hat p},{\hat q};{\bf Q}_{c})$, is
time-dependent through ${\bf Q}_{c}(t)$. We have introduced a
scaled time $s=t/T$ in
order to specify the adiabatic parameter $1/T$ in $\genc$.

If the classical motion is periodic with a period $T$, ${\bf
Q}_{c}(T)={\bf Q}_{c}(0)$, we can perform the adiabatic
expansion with respect to
$(p,q)$ system; since $T$ is usually considered
to be much larger than time scale of
$(p,q)$ system due to the heavy mass.
In general, $h(p,q;{\bf Q}_{c})$ is given by
\eqn\gencc{
h({\hat p},{\hat q};{\bf Q}_{c})\equiv h_{0}({\hat p},{\hat q};{\bf X})+
V({\hat p},{\hat q};{\bf Q}_{c})\quad,
}
where
\eqn\gend{
h_{0}({\hat p},{\hat q};{\bf X})\equiv {1\over 2}\left\{Z(s)%
{\hat p}^2+Y(s)\left({\hat
p}{\hat q}+{\hat q}{\hat p}\right)+X(s){\hat q}^2\right\}\quad,
}
with parameters ${\bf X}\equiv(X(s),Y(s),Z(s))$ being
functions of ${\bf Q}_{c}$. With
this form it is straightforward to estimate the kernel
$K_{{\bf Q}_{c}}(q,q';T)$: first
introduce the source $J$ and the vector notation such that
\eqn\source{
J(s)\equiv\left(\matrix{J_{p}(s) \cr J_{q}(s)\cr}\right)\quad,\quad
\phi(s)\equiv\left(\matrix{p(s)\cr q(s)\cr}\right)
}
and write the potential part in
$\gencc$ as $V(\phi)$, then $\genc$ becomes
\eqn\testaa {
\eqalign{
K_{ {\bf Q}_{c} }(q,q';T) = &\exp\Bigl[-{1\over g}\int_{0}^{1}\!\!ds
\,V\left(g{\d\over \d J} \right)\Bigr]  \cr
&\int\!\!{\cal D} \phi \exp \Bigl[ {1\over g} \int_0^1\!\! ds
\left\{ -{1\over 2} \phi D^{{}^{-1}}\phi + J\phi \right\} \Bigr]
\Bigg\v_{ {J=0}\hfill\atop{
{\bf Q} _{c}(0)={\bf Q} _{c}(T) = {\bf Q} _{c} } }
\cr}
}
where
$$
g\equiv-i{\h\over T}\quad,D^{{}^{-1}}(s,s')\equiv\d
(s,s')\!\!\left(\matrix{Z(s)\quad\quad&Y(s)-{\displaystyle{1\over T}%
{d\over {ds}}}\cr
 Y(s)+{\displaystyle{1\over T}{d\over
{ds}}}&\quad\quad X(s)\cr}\right)\quad.
$$
After integrating out with respect to $\phi$'s, we get
\eqn\testa {
\eqalign{
K_{{\bf Q}_{c}}(q,q';T)= &{\Psi(q,q')\over \sqrt{\det D^{{}^{-1}}}}
\exp\left(-{1\over g}\int_{0}^{1}ds V\left(g{\d\over \d J}\right)\right)\cr
&\times
\exp\left({1\over g}\int_{0}^{1}ds ds' {1\over
2}J(s)D(s,s')%
J(s')\right)\Bigg\v_{J=0},\cr} }
where $D(s,s')$ is the propagator defined by
\eqn\testb {
\int_{0}^{1}ds'D^{{}^{-1}}(s,s')D(s',s'')=\d (s-s'')\quad,
}
and $\Psi(q,q')$ denotes the wave function part (which is irrelevant to
Berry's phase). Inserting the identity;
\eqn\genea{
\exp\left({1\over g}\int_{0}^{1}dsJ(s)\varphi(s)\right)%
\Bigg\v_{\varphi=0}=1\quad,
}
we can rewrite the $J$-derivative in $\testa$ to the $\varphi$-derivative
yielding
\eqn\geneb{
\eqalign{
K_{{\bf Q}_{c}}(q,q';T)=&{1\over \sqrt{\det {
D^{{}^{-1}}}}}\Psi(q,q')\exp\left({g\over 2}\int_{0}^{1}ds ds'{\d\over {\d
\varphi(s)}}D(s,s'){\d\over {\d \varphi(s')}}\right) \cr
&\times \exp\left(-{1\over
g}\int_{0}^{1}dsV(\varphi)\right)\Bigg\v_{\varphi=0}.\cr}
}
{}From this, we can perform a so-called loop-expansion\ref\Col{
S. Coleman, ``Aspects of symmetry," p.135, Cambridge University Press,
Cambridge, 1985.}. The
rule in this case is as follows:  (i)Draw all
possible `bubble-graphs' defined
by no external legs. (ii)Attach $\displaystyle{1\over g}$ to each vertex and
$g$ to each propagator. (iii)Collect graphs having the same power of $g$; a
graph of $g^{\scriptstyle{n-1}}$ is called a n-loop graph.

Since the potential $V(\phi)$ is $O({\phi}^{\scriptstyle{3}})$, bubble graph
starts from the two-loop level, that is, the
leading term is $O(g) = O(1/T)$. On the
other hand, Berry's phase is $O(1)$ quantity (see $\adk$)
which reflects its geometric origin. Thus there is no contribution to
Berry's phase from higher loop graphs; it must be
contained in the determinant $\det
D^{{}^{-1}}$ in $\geneb$.\foot%
{This reminds us of the Adler-Bardeen theorem in
chiral anomaly; which states that chiral
anomaly is determined solely by the
one-loop graph (which is nothing but the determinant part)\ref\Adler{S. L.
Adler and W. Bardeen, Phys. Rev. {\bf 182} (1969),1517.  R. Jackiw, in
``Current
Algebra and Anomalies'' p.137, World Scientific, Singapore, 1985.} .}
{}From this argument, it is
almost clear that the quadratic Hamiltonian $\gend$
which is now called as the generalized harmonic oscillator \ref\Berryc{M. V.
Berry, J. Phys. A:  Math. Gen.  {\bf 18} (1985), 15.  J. H. Hanny, J. Phys.
A: Math. Gen. {\bf 18} (1985), 221.}%
\ref\Jac{R. Jackiw, Int. J. Mod. Phys. {\bf A3}
(1988), 29.} is so fundamental in studying Berry's phase.

The next task is to select out Berry's phase from the $\sqrt{\det {
D^{{}^{-1}}}}$: first recall the expression $\adh$; which
tells us that if we
put $t\mapsto -it$ while
external $\bf X$'s are kept intact, a geometric phase is
unchanged but not for a dynamical phase.
The method of putting $t\mapsto -it$
under path integral is well known as ``Euclidean
technique" \ref\Euca{E. S. Abers and
B.~W. Lee, Phys. Rep. {\bf 9} (1973),
1. T. Kashiwa, Prog. Theor. Phys. {\bf 66} (1981),
1858. T. Kashiwa and M. Sakamoto, Prog. Theor. Phys. {\bf 67} (1982), 1927.}
\ref\Eucc{G. C. Rossi and M. Testa, Ann. Phys.{\bf 148} (1983), 144.}
which is
now briefly reviewed; the starting point is the Feynman kernel,
\eqn\gene {
\eqalign{ {\tilde K}(q,q';T)&\equiv\langle
q|\exp\left(-{T\over \h}H({\hat p},{\hat q})\right)|q'\rangle\cr
&=\lim_{N\ra\infty}\int \prod_{ i=1}^{N}{dp(i) \over {2\pi \h} }
\prod_{j=1}^{N-1} dq(j) \cr  &\times\exp \left( {1\over \h} \sum_{k=1}^{N}
\Bigl\{ ip(k)\D q(k)  -\D tH(p(k),{\bar q}(k)) \Bigr\} \right)
\Biggr\v_{q_{N}=q,q_{0}=q'}, \cr
}}
where we have assumed the time independent Hamiltonian,
for simplicity. By taking
$T\ra\infty$, the first line of $\gene$ becomes
\eqn\genf{
\eqalign{
\lim_{T\ra\infty}{\tilde K}(q,q';T)&=\lim_{T\ra\infty}\sum_{n}%
\langle q|n\rangle
e^{-TE_n/\h}\langle n|q'\rangle\cr
&=e^{-TE_0/\h}\langle q|0\rangle\langle 0|q'\rangle\quad,\cr}
}
where $|n\rangle$'s are eigenstates
of $H({\hat p},{\hat q})$; $H({\hat p},{\hat
q})|n\rangle=E_n|n\rangle$, and $E_{0}$ is the ground state
energy. Thus the path
integral expression of $\genf$ contains the information of
the ground state energy as
well as that of the wave
function, $\psi_{0}(q)\equiv\langle q|0\rangle$\Eucc.
The usual Minkowski representation has no damping factor
thus we cannot reach the
relation like $\genf$: the Feynman kernel is not a
physical quantity which can be
obtained by multiplying wave functions to it. In other
words the knowledge of wave
function is inevitable as in the operator formalism, if we work with the
Minkowski path integral. The Euclidean technique however gives physical
informations to path integral itself.

  According to these, we consider
\eqn\adl{
 -\h{\rd \over \rd \t} \v \P (\t)\ke=H({\hat p},
{\hat q};\bd{X} (\t))\v \P (\t)
\ke \quad  }
instead of the Schr\"odinger
equation $\adb$. Here $\bf{X}(\t)$ is assumed to be
the same function as before; which enables us to use all the relations
$\ada\sim\adaaa$ even in this
case. ( From the discussion, $\genaa\sim\genc$,
it is almost apparent why $\bf{X}$'s,
that is, ${\bf Q}_{c}(t)$ should be left intact
under $t\mapsto -it$; since
otherwise we could get the true ground state of the
heavy mass system, which is out of our interest for the time being.)

Following the same way to $\adk$, we get the Euclidean time
evolution operator $\tilde U (\t)$;
\eqn\adn{
\eqalign{
{\tilde U}(T) =&\lim_{N \ra \infty}\left(1-{1 \over \h}\D
tH(t_{N-1})\right)\times\cdots\times\left(1-{1 \over \h}\D tH(t_{0})\right)
\quad \cr =& \sum_n \exp \left(-{1\over\hbar} \int_{0}^{T}dt E_n(\bd {X} (t)
)+i\gamma_n(T) \right)|n(T)\rangle\langle n(0)| \cr
&\times \Bigl\lbrace 1+O(1/T)
\Bigr\rbrace \quad , \cr}
}
and the Euclidean kernel;
\eqn\ado{
\eqalign{
{\tilde K}(q,q';T) \equiv& \langle q|{\tilde U}(T)|q'\rangle \cr
=& \sum_n \psi_n(q;\bd {X} (T)) \psi_{n}^*(q';\bd {X} (0)) \exp
\left(-{1\over\hbar} \int_{0}^{T}dt E_n(\bd {X} (t) )+i
\gamma_n(T) \right) \cr
&\times\Bigl\{ 1+O(1/T) \Bigr\} \quad . \cr}
}
While the left hand side of $\ado$ has a path integral expression
\eqn\genh{
\eqalign{
{\tilde K}(q,q';T&)
=\lim_{N\ra\infty} \int \prod_{ i=1}^{N}{dp(i) \over {2\pi \h} }
\prod_{j=1}^{N-1} dq(j) \cr
&\times\exp \left( {1\over \h} \sum_{k=1}^{N}
\Bigl\{ ip(k)\D q(k)  - \D tH(p(k),{\bar q}(k);{\bf X}(k)) \Bigr\} \right)
\Biggr\v_{q_{N}=q,q_{0}=q'} \quad.\cr}
}
Hence in view of $\ado$ it should become, by taking $T\ra\infty$, as
\eqn\adp{
\genh \mathop{\sim}_{T\ra\infty}\psi_{0}(q;\bd {X} (T))
\psi_{0}^*(q';\bd {X} (0)) \exp \left(-{1\over\hbar} \int_{0}^{T}dt E_0(\bd
{X}(t) )+i\gamma_0(T) \right) \quad .
}
Note that there is only one imaginary quantity, in the exponent of $\adp$,
which is nothing but a geometric phase.

Hence our proposal for picking up Berry's phase is as follows:
\item{(i)}
Consider the Euclidean kernel for the given Hamiltonian:
$t\mapsto -it$ while
keeping the external variables unchanged.

\item{(ii)} Examine the large T limit
of the kernel.

\item{(iii)} Find the imaginary part of $O(1)$ from the exponent
of the kernel.

These procedures bring us all the informations for
the ground state, including
Berry's phase. As for the excited states,
it is possible to pick up those knowledge
if we start from the kernel subtracting all
contributions of lower levels then
follow the same route as the above.
In order to check these prescriptions, in the
following sections we study simple models.

\newsec{Generalized harmonic oscillator}

Start with the generalized harmonic oscillator discussed in the
previous section
\eqn\osca{
H({\hat p},{\hat q};\bd{X} (t))={1 \over 2}\left\{Z(t){\hat
p}^{2}+Y(t)({\hat p}{\hat q}+{\hat q}{\hat p})+X(t)
{\hat q}^{2}\right\}\quad ,}
with
$X(t)Z(t)-Y^{2}(t) > 0,  Z(t) > 0$ (for the positive definiteness of
$H(\bd{X} (t))$). The Euclidean kernel
$\ado$, is now expressed in view
of $\ctk$ as
\eqn\oscb{
\eqalign{
&{\tilde K}(q,q';T)\equiv \langle q\v {\bf
T}\exp\left(-{T\over\hbar}\int^1_0 ds H({\hat p},{\hat q};{\bf
X}(s))\right)\v q'\rangle =\int{\cal D}p{\cal D}q   \cr
& \times \exp\left({T\over \hbar}\int^1_0 ds\left\{{i\over T}%
p{dq\over ds}-{1
\over 2}\left[Z(s)p^{2}+2Y(s)pq+X(s)q^{2}\right]
\right\}\right)\Biggr\v_{q(0)=q', q(1)=q}. \cr}  }
In the following our calculations are made in the continuum
representation
(but a rigorous treatment can be found in the appendix).
Making a change of
variables such that\foot{Do not confuse $(P, Q)$
with the previous slow
variables.}
\eqn\oscc{
\left(\matrix{q(s)\cr p(s)\cr}\right)
\longmapsto \left(\matrix{Q(s)\cr P(s)\cr}\right)
 = \left(\matrix{\displaystyle{q(s)\over{\sqrt {Z(s)}}}\cr
\sqrt{Z(s)} p(s)+\displaystyle{\alpha(s)q(s)\over\sqrt{Z(s)}} \cr} \right)
}
with
\eqn\oscd{
\alpha(s)\equiv
Y(s)-{i\over 2T}{\dot Z(s)\over Z(s)}\quad ,
}
we obtain
\eqn\osch{
 \eqalign{
{\tilde K} (q,q';T) & = \Phi(q,q')   \cr
 &\times \int{\cal D}P{\cal
D}Q\exp\Bigl[{T\over\hbar}\int_0^1ds\left\{{i\over T}P{dQ\over
ds}-{1\over2}(P^2+{\Omega}^2(s)Q^2)\right\}\Bigr] ,
}}
where
\eqn\osci{
\Phi(q,q')\equiv\exp\left(-{i\over2\hbar}\left\{{\alpha(1)\over
Z(1)}q^2-{\alpha(0)\over Z(0)}{q'}^2\right\}\right)\quad,
}
\eqn\ome{
{\Omega}^2(s)\equiv \omega^2(s)-{1\over T}\o_{1}(s) \quad,
}
with
\eqn\oscj{
\omega^2(s)=X(s)Z(s)-Y^2(s),
\quad \o_{1}(s)=-iZ(s){d \over ds}\left({\a \over
Z}\right)(s)\quad,
}
and other quantity such as a surface term coming from the
Jacobian has been absorbed into
the measure ${\cal D}P{\cal D}Q$.
The $P$-integration is easily performed to give
\eqn\osck{
{\tilde K}
(q,q';T)=\Phi(q,q')\int{\cal
D}Q\exp\left({T\over\hbar}\int_0^1ds\left\{-{1\over2T^2}\left({dQ\over
ds}\right)^2-{1\over2}{\Omega}^2(s)Q^2\right\}\right) \quad .
}
Expanding $Q(s)$ around the classical solution $Q_c(s)$,
\eqn\oscl{
\left(-{d^{2}\over ds^{2}} +
T^{2}{\O}^{2}(s)\right)Q_{c}(s)=0\quad,
}
\eqn\oscla{
Q_{c}(0)={q' \over {\sqrt{
Z(0)}}},\ \ \ Q_{c}(1)={q \over {\sqrt{ Z(1)}}} \quad,
}
such that $Q(s)=Q_{c}(s)+\xi(s)$, we find
\eqn\oscm{
{\tilde K}
(q,q';T)={1\over {\sqrt {2\pi\h\Delta (T)}}}\exp{-1\over \h}S_{c}(q,q';T)
\quad,
}
where
\eqn\oscn{
\eqalign{
S_{c}(q,q';T)&\equiv {i\over 2}
\left\{ {\a(1)\over Z(1)}q^{2}-{\a(0)\over Z(0)}q'^{2}\right\}
+{1\over 2T}\Big[Q_{c}{dQ_{c}\over ds}\Big]_{0}^{1}\quad,\cr
{1\over {\sqrt {2\pi\h\Delta (T)}}} &= \int{\cal D}\xi\exp\left%
(-{T\over
\h}\int_0^1ds{1\over 2}\xi \left(-{1\over T^2}{d^{2}\over
ds^{2}}+{\Omega}^{2}(s)\right)\xi\right),\cr} }
with $\xi$ being satisfied; $\quad\xi (0)=\xi (1)=0\quad .$

According to the standard arguments on the
WKB-approximation\ref\Schulman{L. S. Schulman, ``Techniques and
Applications of Path Integration," p.79, John Wiley \& Sons, New York, 1981.},
the only task needed is to find
out the classical solution $Q_{c}$ in $\oscl$; since
the functional determinant $\Delta (T)$ in $\oscn$, is given by
\eqn\oscnn{
{\displaystyle{1\over {\D (T)}}}={\displaystyle{{{\rd}^{2}S_{c}(q,q';T)}\over
{\rd q \rd q'}}}={1\over 2T}{\displaystyle{{{\rd}^{2}}\over {\rd
q\rd q'}}}\left(\Big[Q_{c} {\displaystyle{{dQ_{c}}\over
{ds}}}\Big]_{0}^{1}\right)\quad.
}
To this end we put the ansatz
\eqn\ansa{
Q_{c}(s)=A\exp\left(T\int_{0}^{s}d\sigma \b(\sigma)\right);\quad
\b(s)=\sum_{\scriptstyle n=0}^{\infty}\b_{n}(s)\left({1\over
T}\right)^{n}\quad,
}
where $A$ is some constant. Plugging $\ansa$ into $\oscl$, we find
\eqn\ansc{
T^{2}\left(\b^{2}(s)-\Omega^{2}(s)\right)+T{\dot \b}(s)=0\quad,
}
which becomes
\eqn\ansd{
\eqalign{
\b^{2}_{0}-\o^{2}&=0;\quad{\rm for}\quad O(T^{2})\quad, \cr
2\b_{1}\b_{0}+{\dot \b}_{0}-\o_{1}
&=0;\quad{\rm for}\quad O(T)\quad,\cr
2\b_{n}\b_{0}+{\dot \b}_{n-1}+\sum_{\scriptstyle m=1}^{n-1}
\b_{n-m}\b_{m}  &=0\quad n \geq 2; \quad{\rm for}\quad
O(T^{-l}) \quad l\geq 0.\cr}
}
For the present purpose, we only need terms
of $O(T^{2})$ and $O(T)$ which are
found to be
\eqn\anse{
\b_{0}(s)=\pm\o(s),\quad
\b_{1}(s)=-{1\over 2}{\dot{\rm \omega }(s) \over \omega
(s)}+{\displaystyle{{\o_{1}(s)}\over {2\b_{0}(s)}}}\quad.
}
Taking into account the boundary condition $\oscla$ we finally get
\eqn\osco{
\eqalign{
Q_c (s) =  {1\over {\sqrt {\omega (s)}} \sinh{\Theta}(1)}
& \left\{ {\sqrt{\omega (1) \over Z(1)}} q \sinh{\Theta}(s)+
 {\sqrt{\omega (0) \over Z(0)}} q'\sinh{\bar\Theta}(s) \right\} \cr
  & \times \Bigl\{1+ O(1/T) \Bigr\} \cr
}}
where
\eqn\defthe{
\eqalign{
\Th (s)&=T\int_{0}^{s}d\sigma \left(\o(\sigma)+{1\over
{2T}}{{\o_{1}(\sigma)}\over {\o(\sigma)}}\right)\cr  &=T\int_{0}^{s}
d\sigma\left\{\o (\sigma) - {i\over {2T}}{Z(\sigma)\over
{\o (\sigma)}}{d\over
{d\sigma}}\left({Y(\sigma)\over Z(\sigma)}\right)\right\} \cr}
}
and ${\bar \Th} (s) = \Th (1) - \Th(s).$  From $\osco$, we thus find
\eqn\oscq{
\eqalign{
{1\over {2T}} &\Big[Q_c{d\over ds}Q_c\Big]_0^1 = {1\over
2\sinh{\Th}(1)}\sqrt{\o(1)\o(0) \over Z(1)Z(0)}\cr
&\times\left\{\left(\sqrt{Z(0)\o(1)
\over Z(1)\o(0)}q^{2} +\sqrt{Z(1)\o(0)\over
Z(0)\o(1)}q'^{2}\right)\cosh{\Th}(1) -2qq'\right\}\quad ,\cr}
}
yielding the determinant $\Delta (T)$ $\oscnn$ to
\eqn\oscp{
{1\over {\sqrt {\Delta (T)}}} = {1\over {\sqrt {\sinh{\Th (T)}}}}%
{\left({{\o (0)\o
(T)}\over {Z(0)Z(T)}}\right)}^{\textstyle{1\over 4}}\quad .
}
Substituting $\oscp$ and $\oscn$ (with $\oscq$) into $\oscm$, we obtain
\eqn\oscr{
\eqalign{
{\tilde K}(q,q';T)=&{1\over \sqrt{2\pi\h\sinh{\Th}(T)}}
\left({\o(T)\o(0) \over Z(T)Z(0)}\right)^{\textstyle{1\over 4}}
\exp\biggl(
{-1 \over 2\h\sinh{\Th}(T)}
\sqrt{\o(T)\o(0) \over Z(T)Z(0)}\cr
&\times \Big\{\Big(
\sqrt{Z(0)\o(T) \over Z(T)\o(0)}q^{2}
+\sqrt{Z(T)\o(0)\over Z(0)\o(T)}q'^{2}\Big)\cosh{\Th}(T)
-2qq'\Big\}\cr
&-{i \over 2\h}\Big({Y(T) \over Z(T)}q^{2}-{Y(0) \over
Z(0)}q'^{2}\Big)\biggr)\quad , \cr}
}
with the original time scale.

Now the informations on the ground state can be derived by putting
$T\ra\infty$;
\eqn\oscs{
\eqalign{
{\tilde K}(q,q';&T)\mathop{\sim}_{T\ra\infty}
\left({\o(T)\over\pi\h Z(T)}\right)^{{\textstyle{1\over 4}}}
\left({\o(0)\over\pi\h Z(0)}\right)^{{\textstyle{1\over 4}}}e^{-{1\over2}%
{\Th}(T)}\cr
&\times\exp\left( -{1\over 2\h Z(T)}\left(\o(T)+iY(T)\right)q^{2}
-{1\over 2\h Z(0)}\left(\o(0)-iY(0)\right)q'^{2}\right)\quad.\cr}
}
Here the imaginary part of $\Th(T)$, $\defthe$,
\eqn\oscv{
\g_{0} (T) = {1\over 4}\int_{0}^{T}dt{Z(t)\over \o(t)}{d \over dt}
\left(Y(t)\over Z(t)\right)
={1\over 4}\int_{\bd{X} (0)}^{\bd{X} (T)}dX^a{Z\over {\sqrt
{XZ-Y^2}}}{\rd\over {\rd X^a}}\left(Y\over Z\right)\quad ,
}
corresponds to the Berry's phase\Berryc\Jac and the real part to the dynamical
one, that is, the (adiabatic) ground state energy,
\eqn\oscu{
E_{0} (\bd{X} (t)) = {\h\over 2}\o (\bd{X} (t))\quad.
}
While $q$ and $q'$ dependent part represents the ground state eigenfunction,
 \eqn\osct{
\psi_{0} (q;\bd{X} (t)) =\left({\o(t)
\over\pi\h Z(t)}\right)^{{\textstyle{1\over
4}}}%
\exp\left( -{1\over 2\h Z(t)}\Bigl(\o(t)+iY(t)\Bigr)q^{2}\right)\quad,
}
yielding
\eqn\oscz{
{\tilde K}(q,q'; T) \mathop{\sim}_{T\ra\infty}\psi_{0}(q;\bd{X}
(T))\psi_{0}^{*}(q';\bd{X} (0)) \exp\Bigl[-{1\over\hbar}\int_{0}^{T}dt E_{0}
(\bd{X} (t))+i\g_{0} (T)\Bigr] ,
}
as it should be.

\newsec{Spin model}

In this section we study another example to convince that our recipe
does also work well in the other path integral representation: the starting
Hamiltonian is
\eqn\spa{
H({\bf{a}}^{\dag},{\bf{a}};{\bd{B} (t)} ) = -\h B(t){\bf{a}}^{\dag}\left(
\bd{n} (t)\cdot \bd{s} \right){\bf{a}}\quad ,
}
where $\bd B (t)$ denotes an external magnetic field,
\eqn\spBa{
\bd B (t)\equiv B(t)\bd{n} (t) =
B(t)\left(\sin\th(t)\cos\phi(t),\sin\th(t)\sin\phi(t),\cos\th(t)\right);
\quad B(t)>0\quad,
}
$$
s_{\a}={\displaystyle{\sigma_{\a}\over 2}}\ ;\quad (\a=1\sim3) \quad,
$$
with $\sigma_{\a}$  being the Pauli matrices and the repeated
induces imply summation.  The symbol ${\bf{a}}$ denotes a two-component
operator, $a_i, (i=1,2)$, with
\eqn\spb{
[a_{i},a^{\dag}_{j}] = \d_{ij} ,
  \quad [a_{i},a_{j}] = [a^{\dag}_{i},a^{\dag}_{j}] = 0\quad
(i,j=1,2)\quad .
}
The basis vectors of Hilbert space is given by
\eqn\spc{
\v m,n\rangle = {1\over {\sqrt{m!n!}}}(a^{\dag}_{1})^m(a^{\dag}_{2})^n\v
0,0\rangle  ,\quad (m,n=0,1,2,\cdots)
}
where $\v 0,0\rangle$ is the vacuum; $ a_i\v 0,0\rangle = 0, (i
=1,2). $ In terms of $\bf {a}$'s the angular momentum operators, $J_{\a}
$'s , and spin amplitude operator, $S$,
are expressed by \ref\Sak{J. J. Sakurai,
``Modern Quantum Mechanics," p.217, Benjamin, 1985.};
\eqn\spd{
J_{\a} = {\h\over 2}{\bf{a}}^{\dag}\sigma_{\a}{\bf{a}} ,\quad S = {1\over
2}{\bf{a}}^{\dag}{\bf{a}}\quad,
}
with
\eqn\spe{
[J_{\a} , J_{\beta}] = i\h\epsilon_{\a\beta\g}
J_{\g} ;\quad [J_{\a} , S] = 0 ;\quad  {\bf{J}}^2 =
\h^2S(S+1)\quad .
}
Due to $J_{\a}$'s and $S$ the infinite set
of $\v m,n\rangle$'s can be classified
into finite dimensional subspaces;
\eqn\spf{
\left\{\v m,n\rangle : m,n = 0,1,2,\cdots\right\}%
 =\left\{\v J,M\rangle : -J\leq M\leq
J,J = 0,{1\over 2},1,{3\over 2},\cdots\right\}\quad ,
}
where $J(\h M)$ is an eigenvalue of $S(J_{3})$. Thus
$H({\bf{a}}^{\dag},{\bf{a}};{\bd{B} (t)})$, $\spa$, can be rewritten as
\eqn\spaa{
H=-{\bf B}(t)\cdot{\bf J}
}
which is nothing but a Hamiltonian describing a spin system coupled with
an external magnetic field[\Bea].
The spin $J$ sector can be picked out by use
of the projection operator $P_J$,
\eqn\spg{
P_{J}=\int_{0}^{2\pi}{d\ld\over 2\pi}\exp
            \left(i\ld({\bf{a}}^{\dag}{\bf{a}}-2J)\right)\quad .
}
To build up a path integral expression, we adopt the coherent state,
$\v{\bf{z}}\rangle$;
\eqn\sph{
\v {\bf{z}}\rangle \equiv \exp ({\bf{a}}^{\dag}{\bf{z}})\v 0,0\rangle, \}uad
{\bf{z}}\ \in\ {\bf{C}}^2\quad,
}
\eqn\spi{
a_i\v {\bf{z}}\ryngle = z_i\v {\bf{z}} \rangle, (
\quad \lengle {\bf{z}}\v {\b~{z}}'\rangle =
\exp({\bf{z}}^{_dag}{\bf{z}}')\quad,
}
\eqn\spk{
\int,!{d{\bf{z}}^{\dag} d{\bf{z}}}\over {{\pi}^{2}}} \v {\bf{z}}\rangle
e^{-{\bf{z}}^{\dag}{\bf{z}}}\langle {\bf{z}} \v ={\bf{1}}\quad .
}
With the aid of the resolution of unity $\spk$, a path integral formula
for this system reads as
\eqn\spl{
\eqalign{
{\tilde K}({\bf {z}},{\bf {z}}';&T)\cr   &\equiv \lim_{N\ra\infty}
\langle { \bf {z}} \Big(1-{{\D t}\over \h}H%
({\bf {a}}^{\dag},{\bf {a} }; {\bf {B}}
(t_{N-1})  ) \Big) \cdots
\Big(1-{{\D t}\over \h}H({\bf {a}}^{\dag},{\bf {a}};{\bf {B} }
(t_{0}) )\Big)
\rangle { \bf{z'} } \cr
&=\lim_{N\ra\infty}\int\prod_{i=1}^{N-1}
     {{d{\bf z}^{\dag}(i)d{\bf z}(i)}\over
{\pi^{2}}}\exp\biggl(-\sum_{k=1}^{N}\Big\{{\bf z}^{\dag}(k)
\left({\bf z}(k)-{\bf z}(k-1)\right)\cr
&-\D t{\bf z}^{\dag}(k)\left({\bf B}(k-1)\cdot{\bf s}\right){\bf
z}(k-1)\Big\}+{\bf z}^{\dag}(N){\bf z}(N)\biggr)
\Bigg\v_{\scriptstyle{\bf z}^{\dag}(N)={\bf
z}^{\dag}\atop\scriptstyle{\bf z}(0)={\bf z}'}\cr}
}
where
${\bf B}(k)\equiv{\bf B}(t_{k})$. Introducing
\eqn\spinm{
{\eta }_{i}\equiv {\bf z}(i) ,\quad {M}_{i,j} \equiv
{\delta }_{i,j}{\bf 1}- \Bigl({\bf 1}+\Delta t{\bf \Omega}(j)
\Bigr){\delta}_{i,j+1},      \quad(1\le i,j\le N-1)\quad,
}
with
\eqn\spome{
{\bf \O}(k)\equiv{\bf B}(k)\cdot{\bf s}=B(k){\bf n}(k)\cdot{\bf s}\quad,
}
we write $\spl$ as
\eqn\spn{
{\tilde K}({\bf z},{\bf
z}';T)=\lim_{N\ra\infty}\int{\displaystyle%
{{d{\bf \eta}^{\dag}d{\bf \eta}}\over
{{\pi}^{2(N-1)}}}}\exp\left(-{\bf \eta}^{\dag}%
M{\bf \eta}+{\bf \eta}^{\dag}{\bf
\zeta}+{\bf \zeta}^{\dag}{\bf \eta}\right) \quad,
}
where
\eqn\spo{
\zeta^{\dag}\equiv\biggl(0,\cdots,0,{\bf z}^{\dag}\left({\bf 1}+\D t{\bf
\O}%
(N-1)\right)\biggr), \quad \zeta\equiv\left(\matrix{\left({\bf 1}+\D t{\bf
\O}(0)\right){\bf z}'\cr 0\cr\vdots\cr 0\cr}\right)\quad.
}
With the aid of the standard technique, that is, a shift such that
\eqn\zzz{
\left(\matrix{\eta\cr \eta^{\dag}\cr}\right)\longmapsto
\left(\matrix{\eta+M^{-1}\zeta\cr
\eta^{\dag}+\zeta^{\dag}M^{-1}\cr}\right)
}
and the integration with respect to $\eta^{\dag}$ and $\eta$ giving the
$\det M$ (whose value is unity in view of $\spinm$), $\spn$ becomes
\eqn\spp{
\eqalign{
{\tilde K}({\bf z},{\bf
z}';T)&=\lim_{N\ra\infty}\exp\left%
(\zeta^{\dag}M^{-1}\zeta\right)\cr
&=\lim_{N\ra\infty}\exp\Biggl({\bf z}^{\dag}\Bigl({\bf 1}+\D t{\bf
\O}(N-1)\Bigr)\cdots \Bigl({\bf 1}+\D t{\bf \O}(0)\Bigr)%
{\bf z}'\Biggr)\quad, \cr }
 }
where $M^{-1}$ is the inverse of $M$. Thus our final task is to evaluate
\eqn\spq{
\eqalign{
R(T)&\equiv\lim_{N\ra\infty}\Bigl({\bf 1}+\D t{\bf \O}(N-1)\Bigr)
\cdots\Bigl({\bf
1}+\D t{\bf \O}(0)\Bigr)\cr
&=\lim_{N\ra\infty}\Bigl({\bf 1}+T\D\t{\bf \O}(N-1)\Bigr)\cdots\Bigl({\bf
1}+T\D\t{\bf \O}(0)\Bigr)\quad, \cr}
}
where the scaled time $\t$ has been
adopted to specify the adiabatic expansion.
It is easy to see that the matrix,
\eqn\spr{
V(k)\equiv e^{i\theta(k)s_{\scriptscriptstyle
2}}e^{i\varphi(k)s_{\scriptscriptstyle 3}} \quad,
}
diagonalizes ${\bf \O}(k)$ such that
\eqn\sprr{
V(k){\bf \O}(k)V^{-1}(k)=B(k)s_{\scriptscriptstyle 3}\quad.
}
Inserting $V^{-1}(k)V(k)$ into $\spq$ successively, we get
\eqn\spRa{
\eqalign{
R(T)=&\lim_{N\ra\infty}V^{-1}(N)V(N)\Bigl({\bf
1}+T\D\t{\bf\O}(N-1)\Bigr)V^{-1}(N-1)\cr
& \times V(N-1) \times\cdots\times
V^{-1}(1)V(1)\Bigl({\bf 1}+T\D\t{\bf \O}(0)\Bigr)V^{-1}(0)V(0)\cr
=&\lim_{N\ra\infty}V^{-1}(N)\Bigl({\bf
1}+T\D\t{\bf \O}^{(1)}(N-1)\Bigr)\cdots\Bigl({\bf 1}+T\D\t{\bf
\O}^{(1)}(0)\Bigr)V(0)\quad,\cr}
}
where
\eqn\spomea{
\eqalign{
{\bf \O}^{(1)}(k)&\equiv B(k)s_{\scriptscriptstyle 3} -
{1\over T}V(k){\dot V}^{-1}(k)\cr
&\equiv{\bf B}^{(1)}(k)\cdot{\bf s}=B^{(1)}(k){\bf n}^{(1)}(k)\cdot{\bf s}
\quad.\cr}
}
Inspecting the component of ${\bf B}^{(1)}$,
\eqn\spBb{
\eqalign{
B^{(1)}_{\a}&=2\Tr\left[s_{\a}s_{\scriptscriptstyle 3}
B-{1\over T}s_{\a}V{\dot V}^{-1}\right]\cr
&=\left(-{i\over T}{\dot \varphi}\sin\theta,
{{i{\dot  \theta}}\over T}, B+{i\over T}{\dot
\varphi}\cos\theta\right)\quad, \cr}
}
and comparing it with $\spBa$, we
see that the first and the second component
of the vector ${\bf B}^{(1)}$ reduce to $O(1/T)$. Going on with the same
manner, we find
\eqn\spRb{
\eqalign{
R(T)=\lim_{N\ra\infty}&V^{-1}(N)V^{(1)-1}(N)\cdots
V^{(j-1)-1}(N)\left({\bf
1}+T\D\t{\bf \O}^{(j)}(N-1)\right)\times\cdots\cr
&\times\left({\bf 1}+T\D\t{\bf \O}^{(j)}(0)\right)V^{(j-1)}%
(0)\cdots
V^{(1)}(0)V(0)\quad, \cr}
 }
where
\eqn\spBc{
\eqalign{
V^{(j)}(k)&\equiv e^{i\theta^{(j)}(k)s_{\scriptscriptstyle 2}}
e^{i\varphi^{(j)}(k)s_{\scriptscriptstyle 3}}\quad,\cr
{\bf \O}^{(j)}(k) &\equiv B^{(j-1)}(k)s_{\scriptscriptstyle 3}-{1\over
T}V^{(j-1)}(k){\dot V}^{(j-1)-1}(k)\cr &\equiv{\bf B}^{(j)}(k)\cdot{\bf
s}=B^{(j)}(k){\bf n}^{(j)}(k)\cdot{\bf s}\quad,\cr}
}
with
\eqn\spBd{
\eqalign{
&{\bf B}^{(j)}\equiv\left(-{i\over T}%
{\dot \varphi}^{(j-1)}\sin\theta^{(j-1)},
\quad{{i{\dot  \theta}^{(j-1)}}\over T}, \quad B^{(j-1)}+{i\over
T}{\dot\varphi}^{(j-1)}\cos\theta^{(j-1)}\right) \quad, \cr
&\tan\theta^{(j)}\equiv{\displaystyle{{-{\displaystyle{i\over
T}}{\sqrt{\left({\dot \varphi}^{(j-1)}
\sin\theta^{(j-1)}\right)^{2}+\left({\dot \theta}^{(j-1)}%
\right)^{2}}} \over
{B^{(j-1)}+{\displaystyle{i\over
T}}{\dot\varphi}^{(j-1)}\cos\theta^{(j-1)}}}}} \quad, \cr
&\tan\varphi^{(j)}\equiv-{\displaystyle%
{{{\dot \theta}^{(j-1)}}\over {{\dot
\varphi}^{(j-1)}\sin\theta^{(j-1)}}}} \quad.\cr}
 }
{}From $\spBd$ and $\spBb$, $O(\theta^{(j)})
=O(T^{-j})$, thus up to $O(T^{-j})$
we have succeeded in diagonalizing the
original ${\bf \O}(k)$ to yield ${\bf
\O}^{(j)}(k)$.

For our purpose it is enough to perform two successive
transformations; since the Hamiltonian in each factor
of $R(T)$ is $T{\bf \O}$
then taking into all the terms of
$O(T^{-1})$ is sufficient. To this order, we
have from $\spBc$ with $\spBb$
\eqn\omethi{
{\bf \O}^{(2)}=B^{(2)}s_{\scriptscriptstyle 3} + O\left%
({ \displaystyle{1\over T^{2}}
}\right)  \quad,
}
with
\eqn\spBe{
\eqalign{
B^{(2)}&=B+{i\over T}\left({\dot\varphi}\cos\theta+{\dot \varphi}^{(1)}
\right)+O\left({ \displaystyle{1\over T^{2}} }\right)\cr
 \varphi^{(1)}&=-\tan^{-1}\left({ \displaystyle{ {\dot\theta}\over{
{\dot\varphi}\sin\theta} } }\right) \quad.\cr}
}
Also from $\spr$ or $\spBc$
\eqn\spVa{
V^{(1)}V=\left(
\matrix{e^{{i\over2}(\varphi+\varphi^{(1)})}\,\cos{\displaystyle{\theta\over
2}} &\quad\quad
e^{-{i\over2}(\varphi-\varphi^{(1)})}\,
\sin{\displaystyle{\theta\over 2}}&\cr
-e^{{i\over2}(\varphi-\varphi^{(1)})}\,\sin{\displaystyle{\theta\over 2}}&
e^{-{i\over2}(\varphi+\varphi^{(1)})}\,
\cos{\displaystyle{\theta\over2}}&\cr}\right)+O({\displaystyle{1\over T}})
}
by noting that $\theta^{(1)}=O(1/T)$. Plugging $\spBe$ (with $\omethi$) and
$\spVa$ into $\spRb$, we finally get after a bit of long calculation;
\eqn\spRR{
 \eqalign{
R(T) = \left(
\matrix{
e^{\Th (T)}\cos \displaystyle{\th \over 2}\cos {\th' \over 2}
 & \quad e^{\Th (T)-i\phi'}\cos \displaystyle%
{\th \over 2}\sin \displaystyle{\th'
\over 2}\cr +e^{-\Th (T)-i(\phi-\phi')}\sin \displaystyle{\th \over 2}\sin
\displaystyle{\th' \over 2} &-e^{-\Th (T)-i\phi}\sin \displaystyle{\th \over
2}\cos \displaystyle{\th' \over 2}\cr & \cr
& \cr
 e^{\Th (T)+i\phi}\sin \displaystyle%
{\th \over 2}\cos \displaystyle{\th' \over 2}
 & e^{\Th (T)+i(\phi-\phi')}\sin \displaystyle{\th \over 2}\sin
\displaystyle{\th' \over 2}\cr -e^{-\Th (T)+i\phi'}%
\cos \displaystyle{\th \over
2}\sin \displaystyle{\th' \over 2} &+e^{-\Th (T)}%
\cos \displaystyle{\th \over
2}\cos \displaystyle{\th' \over 2}\cr}\right) \quad .\cr}
}
where
\eqn\sps{
\Th (T) = {T\over 2}\int_{0}^{1}d\t\left\{B(\t)-{i\over T}%
\left(1-\cos{\th
(\t)}\right){{d\phi}\over {d\t}}(\t)\right\}\quad .
}

The $P_J$-inserted kernel ${\tilde K}^{(J)}({\bf{z}},{\bf{z}}';T)$ is also
found in a quite parallel manner; first noting that
\eqn\xxx{
e^{i\ld {\bf a}^{\dag} {\bf a} } \v {\bf z}
\rangle=\v e^{i\ld}{\bf z} \rangle
 }
we get
\eqn\spt{
\eqalign{
&{\tilde K}^{(J)}({\bf{z}},{{\bf{z}}'};T) \cr
      &  =\lim_{N\ra\infty}\int_{0}^{2\pi}{d\ld \over 2\pi}
          \langle {\bf z} \v
          \left(1-{\D t \over \h}H(t_{N-1})\right)\cdots
          \left(1-{\D t \over \h}H(t_{0})\right)
          e^{i\ld ({\bf a}^{\dag} {\bf a} -2J })
          \v {\bf z'} \rangle\cr
         &  =\int_{0}^{2\pi}{d\ld\over 2\pi}
          {\tilde K} ({\bf z} , e^{i\ld} {\bf z'} ;T)e^{-2iJ\ld}
         =\int_{0}^{2\pi}{d\ld\over 2\pi}
          \exp\Big({\bf z}^{\dag} R(T) {\bf z'} e^{i\ld} -2iJ\ld\Big)\cr
    &={1\over (2J)!}\left(\bd{z} ^{\dag} R(T) \bd{z'} \right)^{2J},
\cr}
}
where we have used the previous result $\spp$ with $\spq$ in the second line.
We thus find
\eqn\spu {
{\tilde K}({\bf{z}},{{\bf{z}}'};T) = \sum_{\scriptstyle 2J=0}^\infty
{1\over {(2J)!}}\left({\bf{z}}^{\dag}R(T){{\bf{z}}'}\right)^{2J}\quad.
}
Although the spectrum of the total Hamiltonian $\spa$ extends from
$-\infty$ to $+\infty$, which prevents us from putting $T$ infinity
in ${\tilde K}({\bf{z}},{{\bf{z}}'};T)$, that of ${\tilde
K}^{(J)}({\bf{z}},{{\bf{z}}'};T)$ does not;
\eqn\spv{
\eqalign{
{\tilde K}^{(J)}({\bf {z}},{{\bf {z}}'}; T) \cr
\mathop{\sim}_{T\ra \infty} e^{2J\Th(T)}{1\over {(2J)!}}
\left(z^{*}_{1}\cos{\th\over 2}+z^{*}_{2}e^{i\phi}\sin{\th\over 2}
\right)^{2J} \left(z'_{1}\cos{\th'\over 2}+z'_{2}e^{-i\phi'}\sin{\th'\over 2}
\right)^{2J}\cr
 =\psi^{J}_{0}({\bf{z}};{\bd{B} (T)}){\psi^{J}_{0}}^{*}({{\bf{z}}'};{\bd{B}
(0)})
\exp\Bigl[JT \int_{0}^{1}d\t\left\{B(\t)-{i\over
T}\left(1-\cos{\th(\t)}\right){{d\phi}\over {ds}}(\t)\right\}\Bigr] ,\cr
} }
where by use of $\spRR$ $\psi^{J}_{0}({\bf{z}};{\bd{B} (t)})$
denotes the instantaneous eigenfunction of the ground
state for spin $J$ sector
\eqn\spx{
\eqalign{
\psi^{J}_{0}({\bf{z}};{\bd{B}
(t)})=  & {1\over\sqrt{(2J)!}}\left(z^{*}_{1}\cos{\th(t)\over
2}+z^{*}_{2}e^{i\phi(t)}\sin{\th(t)\over 2} \right)^{2J}\cr
= & \sum_{\scriptstyle M=-J}^{\scriptstyle J}\left[{{(2J)!}\over
 {(J+M)!(J-M)!}}\right]^{1\over 2}\left(\cos{{\th (t)}\over
2}\right)^{J+M}  \cr
& \qquad \qquad \times \left(e^{i\phi (t)}\sin{{\th (t)}\over
2}\right)^{J-M}\langle {\bf{z}} \v J,M\rangle,\cr
}  }
with $\langle {\bf{z}} \v J,M\rangle$ being eigenfunctions of $J_{3}$;
\eqn\spy{
\langle {\bf{z}} \v J,M\rangle={1\over
 {\sqrt{(J+M)!(J-M)!}}}\left(z^{*}_{1}\right)^{J+M}\left(z^{*}_{2}
\right)^{J-M}\quad .
}
In $\spv$, the imaginary part corresponds to the Berry's
phase\Bea,
\eqn\spz{
\eqalign{
\g_{0} &= -J\int_{0}^{T}dt\left
(1-\cos{\th (t)}\right){{d\phi}\over{dt}}(t)\cr
&= -J\int_{\phi'}^{\phi}d\phi \left(1-\cos{\th}\right)\quad ,\cr}
}
as it should be and the real part of the exponent represents the ground
state energy
\eqn\xxa{
E_{0}\left({\bd{B} (t)}\right) = -\h JB(t)\quad .
}

Needless to say that our method of calculations is easily
genelarized to more general cases such as $SU(1,1)$
(which is nothing but the previous case\Jac\ \ref\Hira{M. Hirayama, Toyama
University Preprint, March 1991.}) and others also.

\newsec{Discussions}

In this paper we have discussed the method of
picking up Berry's phase under
the path integral formalism. With the aid of Euclidean technique and the
adiabatic expansion the imaginary part of
$O(1)$ term is identified as Berry's
phase . Models we have adopted are so simple but general:  the generalized
harmonic oscillator of single degree is universal
as for Berry's phase.  In the
spin case the method of coherent representation
is so flexible in the sense that
we can extend it($SU(2)$ group) to a more general group such as $SU(1,1)$
which is nothing but the dynamical group of the generalized harmonic
oscillator[\Jac, \Hira], SU(3), etc.. It
also allows us to give a tractable treatment
in the adiabatic approximation to study
geometrical properties. The situation is
quite controversial in the case of path
integral fomula invented by Nielsen and
Rohrlich\ref\NR{H. B. Nielsen and D. Rohrlich, Nucl. Phys. {\bf B299} (1988),
471. T. Kashiwa, Int. J. Mod. Phys.
{\bf A5} (1989), 375.}; since there is only one
degree of freedom in their spin model and the domain of integration itself
represents a non-trivial topology; punctured sphere. The Berry's phase is
hidden in that topology.  The resultant equations are consequently highly
nonlinear, which makes it subtle to solve the model with a method of unitary
transformation except calculating the path integral exactly.

   We have introduced a systematic expansion with respect to the adiabatic
parameter, as can be recognized from
$\ansa$ $\sim$ $\ansd$ also from $\spq$
$\sim$ $\spBd$,  which enables us to go to any order in the
expansion\ref\Beb{M. V. Berry, Proc. R. Lond. {\bf A414} (1987), 31.}.

 Our assumption that the external variables
should be left intact under the
Euclidean technique comes from the
fact that since otherwise the fiducial level
given by the heavy mass system becomes movable to break down the basic
hypothesis of the adiabatic approximation.
Nevertheless, it is captivating to
study consequences from the whole Euclidean technique
in the light of the
adiabatic approximation
\ref\DR{ W. Dittrich and M. Reuter, Phys. Lett. {\bf
A155} (1991), 94.}.

 So far we have concentrated only on Berry's
phase since in path integral the
adiabatic approximation is nothing but
the WKB-approximation and it is so easy
to carry out. On the other hand, Aharonov-Anandan phase\AA\ which is, from
our point of view, conceptually more significant, is yet unamenable to
calculations. It might thus be
interesting and important to study that in terms
of the path integral language.

\appendix{A}{A rigorous treatment of the path integral $\oscb$}

 In this appendix, we make a more careful
derivation of the result $\oscr$ with
a short review of the path integral
formula $\dtk$: if a Hamiltonian is written
by
\eqn\apa{
{\hat H}(t)=H({\hat p},{\hat q};{\bf X}(t))
}
with the caret representing operators, then
the (Euclidean) time evolution
operator is
\eqn\apb{
{\tilde U}(T)=\lim_{N \ra \infty}\left(1-{T \over\h}\D sH(t_{N-1})\right)
\times\cdots\times\left(1-{T \over \h}\D sH(t_{0})\right)\quad .
}
By applying completeness relations,
\eqn\apac{
\int_{-\infty}^{+\infty}dq\v q\rangle\langle q\v={\bf{1}},\quad\quad
\int_{-\infty}^{+\infty}dp\v p\rangle\langle p\v={\bf{1}}\quad ,
}
successively to $\apb$, the path integral representation of Feynman
kernel is found to be;
\eqn\apc{
\eqalign{{\tilde K}(q,q';T)=\langle q|{\tilde U}(T)|q'\rangle
=\lim_{N\ra\infty}\int \prod_{ i=1}^{N}{dp(i) \over {2\pi \h} }
\prod_{j=1}^{N-1} dq(j)\quad \quad \quad \quad \quad \quad \quad \cr
\ \ \ \ \ \ \ \ \times\exp \left( {1\over \h}
\sum_{k=1}^{N} \Bigl\{ ip(k)\D
q(k)  -\D tH(p(k),{\bar q}(k);{\bf X}(k)) \Bigr\} \right)
\Biggr\v_{q_{N}=q,q_{0}=q'}\quad, \cr}
}
where ${\bf X}(k)\equiv {\bf X}(t_{k})$,
\eqn\apd{
\D q(k)\equiv q(k)-q(k-1), \quad \D t\equiv T/N\quad,
}
\eqn\ape{
{\bar q(k)} \equiv {1\over 2}{(q(k)+q(k-1))}\quad.
}
In $\apc$, we have assumed the ordering of the Hamiltonian $\apa$ to be the
Weyl-ordered form;
\eqn\apg{
 \left\{{\hat p}^{m}{\hat q}^{n}\right\}_{W}\equiv\left({\rd \over \rd
ia}\right)^{m}\left({\rd \over \rd ib}\right)^{n}e^{\displaystyle{ia{\hat
p}+ib{\hat q}}}\v_{a=b=0}
}
thus yielding the mid-point prescription $\ape$. In the case of $\osca$, the
Hamiltonian in $\apc$ is
\eqn\aph{
  H(p(k),{\bar q}(k);{\bf X}(k))={1\over 2}\Bigl\{Z(k)p(k)^2+2Y(k)p(k){\bar
  q}(k)+X(k){\bar q}(k)^2\Bigr\}\quad.
}

        It is rather cumbersome to get the corresponding rule to $\oscc$ and
$\oscd$ in the discrete representation. To this end we follow the method of
canonical transformation\ref\FK{H. Fukutaka and T. Kashiwa, Ann. Phys. {\bf
185} (1988), 301.}: first notice that the transformation,
\eqn\oscc{
\left(\matrix{q(s)\cr p(s)\cr}\right)
 \longmapsto \left(\matrix{Q(s)\cr P(s)\cr}\right)
= \left(\matrix{{\displaystyle{q(s)\over{\sqrt {Z(s)}}}}\cr
\sqrt{Z(s)} p(s)+{\displaystyle{\alpha(s)q(s)\over\sqrt{Z(s)}}} \cr} \right)
\quad,
}
with
\eqn\oscd{
\alpha(s)\equiv
Y(s)-{i\over 2T}{\dot Z(s)\over Z(s)}\quad ,
}
comes from the canonical transformation;
\eqn\api{
ipdq-H(p,q;{\bf X}(t))dt=iPdQ-K(P,Q;{\bf X}(t))dt+idF\quad,
}
with
\eqn\apj{
F\equiv-PQ+{1\over {\sqrt Z}}Pq-{\alpha\over {2Z}}q^{2}\quad.
}
where $i={\sqrt -1}$ is the consequence of
the Euclidean technique. Hence our
starting point should be
\eqn\apk{
 ip(k)\D q(k) -H(p(k),{\bar q}(k);{\bf X}(k))%
\D t=iP(k)\D Q(k) -K(k)\D t+i\D
F(k)
}
with $\apj$. Here
\eqn\apai{
\eqalign{
\D F(k)& \equiv F(k)-F(k-1)\cr
&=-P(k)\D Q(k)-\left\{Q(k-1)-{1\over
{\sqrt {Z(k-1)}}}q(k-1)\right\}\D P(k)\cr
&+\left\{{1\over {\sqrt {Z(k)}}}P(k)-{{\a(k)}\over {Z(k)}}%
{\bar q}(k)\right\}\D
q(k)\cr &+\D \left({1\over {\sqrt Z}}\right)(k)q(k-1)%
P(k)-{1\over 2}\D \left({\a \over
Z}\right)(k)q(k-1)^2\quad .\cr}
}
Inserting $\apai$ into $\apk$ and comparing the both sides with respect $\D
q(k)$ and $\D P(k)$ terms, we get a desired transformation:
\eqn\apae{
\left(\matrix{q(k)\cr p(k)\cr}\right)
 \longmapsto \left(\matrix{Q(k)\cr P(k)\cr}\right)
= \left(\matrix{{\displaystyle{q(k)\over{\sqrt {Z(k)}}}}\cr
 \sqrt{Z(k)} p(k)+{ \displaystyle%
{ \alpha(k)\over\sqrt{Z(k)} } }{\bar q}(k) \cr}
\right)
}
or
\eqn\apl{
\left(\matrix{q(k)\cr p(k)\cr}\right)
=\left(\matrix{{\displaystyle{1\over{\sqrt{Z(k)}}}}\left%
(P(k)+\a (k){\bar
Q}(k)\right)+O(\D t)Q(k-1)\cr
  \sqrt{Z(k)}Q(k) \cr}\right)\quad.
}
$\D t$ terms in both sides also tell us that
\eqn\apm{
\eqalign{
 K(k)=&H\left({\displaystyle{1\over{\sqrt{Z(k)}}}}\left(P(k)+\a (k){\bar
Q}(k)\right),\sqrt{Z(k)}{\bar Q}(k);{\bf X}(k)\right)\cr
 &-i{{{\dot Z}(k)}\over
{2Z(k)}}Q(k-1)P(k)-i{1\over2}Z(k){d\over {dt}}\left({\a
\over Z}\right)(k)Q(k-1)^2\quad,\cr}
}
where the differential symbol for the external variables has been adopted
such that
\eqn\qqz{
{\dot {\bf X}}(k)\equiv{d\over dt}{\bf X}(k)\equiv{1\over
{\D t}}\left({\bf X}(k)-{\bf
X}(k-1)\right) \quad,
}
and we have discarded all the $O((\D t)^2)$ terms
according to the spirit of
path integral. In addition to these, however, we must take into account the
surface terms:
\eqn\apag{
\sum_{\scriptstyle k=1}^{N} \D F(k)=F(N)-F(0)\quad,
}
whose $F(N)$ and $F(0)$ represent the corresponding unitary
transformation to
the wave function.\par

{}From $\apae$ or $\apl$,
\eqn\zazq{
 \prod_{\scriptstyle i=1}^{N}{{dp(i)}\over
 {2\pi\h}}\prod_{\scriptstyle j=1}^{N-1}dq(j)={1\over
{\sqrt{Z(N)}}}\prod_{\scriptstyle
i=1}^{N}{{dP(i)}\over{2\pi\h}}\prod_{\scriptstyle
j=1}^{N-1}dQ(j)\quad.
}
Hence the transformed kernel is
\eqn\apak{
\eqalign{
&{\tilde K}(q,q'; T) = \lim_{N \rightarrow \infty} {1
\over {\sqrt{Z(N)}} } \Phi(q,q') \int \prod_{i=1}^{N}
{{dP(i)}\over 2\pi \hbar} \prod_{j=1}^{N-1}
dQ(j) \cr
& \times \exp\Biggl[ {1 \over \hbar} \sum_{k=1}^{N}
\biggl\{ iP(k)\Delta Q(k) - {\Delta t\over 2} \bigl[
{P(k)}^2 + 2(Y(k) - \alpha (k) ) P(k) \bar Q(k) \cr
& \qquad \quad + \bigl({\omega (k)}^2 + (Y(k)-\alpha
(k))^2\bigr) {{\bar Q}(k)}^2 \bigr] + {{i \Delta t} \over
2} \bigl[ {{\dot Z(k)} \over Z(k)} Q(k-1) P(k) \cr
& \qquad \qquad \qquad \qquad \qquad  + Z(k) {{d\over dt}}
({\alpha \over Z})(k) {Q(k-1)}^2 \bigr] \biggr\} \Biggr]
\cr
}}
where
\eqn\apal{
\Phi (q,q')\equiv\exp\left(-{i\over2\hbar}\left\{{\alpha(N)\over
Z(N)}q^2-{\alpha(0)\over Z(0)}{q'}^2\right\}\right)\quad,
}
which has appeared due to the surface term $\apag$.

It is still necessary to do more; we must convert $Q(k-1)$, in $\apm$,
that is, in the third line in
$\apak$, into the mid-point ${\bar Q(k)}$ $\ape$. To
this end we first note that
\eqn\zxw{
 Q(k-1)={\bar Q}(k)-{1\over 2}\D Q(k)\quad ,
}
and substitute this into the third line of $\apak$; the first term becomes
\eqn\apam{
  {i\over 2}\D t {{{\dot Z}(k)}\over
{Z(k)}}Q(k-1)P(k)={i\over 2}\D t {{{\dot
Z}(k)}\over {Z(k)}}{\bar Q}(k)P(k)-{i\over 4}\D t {{{\dot Z}(k)}\over
{Z(k)}}\D Q(k)P(k)
}
whose first term kills $\left(Y(k)-\a (k)\right)P(k){\bar Q}(k)$ term in the
second line of $\apak$ due to $\oscd$.
The second term of $\apam$ is combined
with the first term in the exponent of $\apak$ to give
\eqn\apn{
{1\over
\h}\sum_{\scriptstyle k=1}^{N}i\left(1-{{{\dot Z}(k)}\over {4Z(k)}}\D
t\right)P(k)\D Q(k)\quad.
}
The shift
\eqn\zxz{
 P(k)\longmapsto
{\displaystyle{P(k)\over{1-{\displaystyle{{{\dot Z}(k)}\over
{4Z(k)}}}\D t }}}\quad,
}
brings us to the usual kinetic term leaving the Jacobian,
\eqn\apo {
\eqalign{
 \prod_{\scriptstyle k=1}^{N}&\left\{1-{{{\dot Z}(k)}\over  {4Z(k)}}\D
t\right\}^{-1}=\exp\left(\D t \sum_{\scriptstyle k=1}^{N}{{{\dot Z}(k)}
\over
{4Z(k)}}\right)\cr &=\exp\left({1\over 4}\int_{0}^{T}dt{d\over
dt}\ln{Z(t)}\right)=\left({{Z(N)}\over {Z(0)}}\right)^ {\textstyle{1\over
4}}\quad,\cr}
}
(where we have kept the term $O(\D t)$ in the first
line and converted the sum
into the integration in the second line.)
With a similar manner, we can have a
substitution;
\eqn\app{
{{i\D t}\over 2}Z(k){d\over {dt}}\left({\a \over
Z}\right)(k)Q(k-1)^2\longmapsto {{i\D t}\over 2}Z(k){d\over {dt}}\left({\a
\over Z}\right)(k){\bar Q}(k)^2\quad.
}
(In this case there is no contribution from a Jacobian.)
Thus we get
\eqn\apap{
\eqalign{
&{\tilde K}(q,q'; T) = \lim_{N \rightarrow \infty}
{\Phi(q,q')\over {Z(N)Z(0)} ^{1\over 4}}  \int
\prod_{i=1}^{N} {{dP(i)}\over 2\pi \hbar}
\prod_{j=1}^{N-1} dQ(j) \qquad \qquad \qquad \cr
& \times \exp\Biggl[ {1 \over
\hbar} \sum_{k=1}^{N} \biggl\{ iP(k)\Delta Q(k) - {\Delta
t\over 2} \bigl[ {P(k)}^2 +  \bigl({\omega (k)}^2 +
(Y(k)-\alpha (k))^2 \cr
&\qquad \qquad \qquad \qquad  -i Z(k){{d\over dt}} ({\alpha
\over Z})(k) \bigr) {{\bar Q}(k)}^2  \bigr] \biggr\}
\Biggr] \cr }
}
under the boundary conditions;
\eqn\apam{
 Q(N)={q\over{\sqrt{Z(N)}}}\quad,\quad
Q(0)={q'\over{\sqrt{Z(0)}}}\quad.
}

Now we estimate $\apap$: first the integration with respect to $P$ gives
\eqn\apaq{
\eqalign{
 {\tilde K}&(q,q';T)=\lim_{N\ra \infty}{1\over
{\left(Z(N)Z(0)\right)^{\textstyle{1\over 4}}\left(2\pi\h\D
t\right)^{\textstyle{N\over 2}}}}\Phi (q,q')\int\prod_{\scriptstyle
j=1}^{N-1}dQ(j)\cr
&\exp\Biggl({1\over\hbar}\sum_{\scriptstyle k=1}^{N} \Biggl\{-{1\over
{2\D t}}\left(\D Q(k)\right)^2 \cr
&-{{\D t}\over {2}}\left(\o (k)^2-\left({{\dot
Z}\over {2TZ}}\right)^2-i{{Z(k)}\over T}{d\over {ds}}\left({\a \over
Z}\right)(k)\right){\bar Q}(k)^2\Biggr\}\Biggr)\quad,\cr
}  }
where we have shifted the time derivative of the external variables
$t\mapsto sT$ to clear up the adiabatic approximation.
We rewrite the exponent as follows:
\eqn\apar{
\eqalign{
&{\rm exponent\ \ of\ \ \apaq}\cr
&=-{1\over 2{\h\D t}}\Bigl\{\left({\bf{Q}}-M_{N}^{-1}{\bf{d}}\right)^{\rm
T}M_{N}\left({\bf{Q}}-M_{N}^{-1}{\bf{d}}\right)-{\bf {d}}^{\rm
T}M_{N}^{-1}{\bf {d}} \cr
& +\left(Q(0)^2+Q(N)^2\right)\Bigr\} +O(\D t)\quad ,\cr
}}
where
\eqn\axz{
 {\bf {Q}}\equiv\left(\matrix{Q(1)\cr Q(2)\cr\vdots\cr Q(N-2)\cr
Q(N-1)\cr}\right)\ \,\quad {\bf {d}}\equiv\left(\matrix{b(0)Q(0)\cr
0\cr\vdots\cr 0\cr b(N-1)Q(N)\cr}\right)\quad ,
}
\eqn\apmat{
M_{N}\equiv\left(\matrix{
               a(1) &-b(1)&{}      &\cr
              -b(1)& a(2)&-b(2) &       & 0 &\cr
                       &-b(2)& a(3) &\ddots &\cr
                           &     &\ddots&\ddots &\cr
                   & 0  &      &       &-b(N-2)\cr
                   &     &      &-b(N-2)&a(N-1)\cr}\right)\quad,
}
with
\eqn\apas{
\eqalign{
&a(k)=2+{{{\D t}^2}\over 2}{\o (k)}^2\left[1-{{2i}\over T}{{Z(k)}
 \over {{\o (k)}^2}}{d\over {ds}}
\left({\a\over Z}\right)(k)-\left({{\dot Z}\over
{2T\o Z}}\right)^{2}(k)\right]\quad,\cr
&b(k)={1-{{{\D t}^2}\over 4}{\o (k)}^2\left[1-\left({{\dot Z}\over {2T\o
Z}}\right)^{2}(k)\right]}\quad .\cr}
}
The Gaussian integration over $Q$ yields
\eqn\apat{
\eqalign{
{\tilde K}(q,q';T)=&\lim_{N\ra\infty}{1\over
{\left(Z(N)Z(0)\right)^{\textstyle{1\over 4}}}\left( 2\pi\h\D
t\right)^{\textstyle{1\over 2}}}
{1\over {\sqrt {{\rm det}M_{N}}}}\Phi (q,q')\cr
&\times\exp\left({1\over {2\h\D t}}\left\{{\bf {d}}^{\rm T}M_{N}^{-1}{\bf
{d}}- \left(Q(0)^2+Q(N)^2\right)\right\}+O(\D t)\right)\quad.\cr}
}

Now we estimate the determinant and $M_{N}^{-1}$: first write
\eqn\apdet{
{\rm det}M_{N}\equiv D(N-1)\quad.
}
With the aid of the Laplace expansion it is shown that $D(N-1)$ satisfies
\eqn\apq{
\eqalign{
D(N-1)=&a(N-1)D(N-2)-b^2(N-2) D(N-3)\cr
&=a(1){\tilde D}(N-2)- b^2(2){\tilde D}(N-3)\ \ \ \ \ \ \ \cr}\quad ,
}
 where the expansion has started from the ($N-1$)-th row and collum so
$D(N-2)$ is the cofactor of
$(M_{N}){_{N-1, N-1}}$ in the first line, and  from
the first row and collum so
${\tilde D}(N-2)$ is that of  $(M_{N}){_{1, 1}}$ in the
second line. By putting
\eqn\apr{
  \Delta sD(k)\equiv f(s),\quad \Delta
s\tilde{D}(k)\equiv f(1-s);\quad s\equiv k\Delta s ,\quad(0\le s\le 1)
\quad,
}
$\apq$ becomes the differential equation \ref\SchulB{L. S. Schulman,
``Techniques and Applications of Path
Integration," p.31, John Wiley \& Sons,
New York, 1981.}
\eqn\aps {
\eqalign{
\left(-{d^{2}\over ds^{2}} +
T^{2}{\tilde \O}^{2}(s)\right)f(s)=0\quad, \cr
f(0)=0,\quad {\dot f}(0)=1  \ \quad, \ \ \ \ \ \cr}
}
with
\eqn\omegat{
{\tilde {\O}}(k)^2=\o
(k)^2-{{iZ(k)}\over T}{d\over {ds}}\left({Y\over Z}\right)(k)-{{2Z(k){\ddot
Z}(k)-3{\dot Z}(k)^2}\over {4T^{2}Z(k)^{2}}}\quad.
}
The following procedure for finding the
solution is quite similar as that from
$\ansa$ to $\osco$ with the replacement ${\O}^2
\mapsto {\tilde \O}^2$, thus
we skip it to give the final answer
\eqn\apt{
f(s)={1 \over T}{1 \over \sqrt {\omega (s)\omega (0)}}\rm sinh\Theta (s)
\quad,
}
with $\Theta$ being given by $\defthe$. By noting
\eqn\imatc{
\eqalign{
\left(M_{N}^{-1}\right)_{\scriptstyle 1,1} &= {{\tilde D}(N-2)
\over D(N-1)} = {f(1-\Delta s) \over f(1)} = 1-{\Delta s{\dot f}(1)
\over f(1)} \quad,\cr
\left(M_{N}^{-1}\right)_{\scriptstyle N-1,N-1}&={D(N-2) \over
D(N-1)} = {f(1-\Delta s) \over f(1)} =1-{\Delta s{\dot f}(1) \over
f(1)}\quad,\cr
\left(M_{N}^{-1}\right)_{\scriptstyle
1,N-1}&=\left(M_{N}^{-1}\right)_{\scriptstyle N-1,1} = {1\over
D(N-1)} = {{\Delta s} \over f(1)} \quad, \cr}
}
and using $\apt$, it is apparent that the kernel $\apat$ becomes $\oscr$;
\eqn\oscix{
\eqalign{
{\tilde K}(q,q';T)=&{1\over \sqrt{2\pi\h\sinh{\Th}(T)}}
\left({\o(T)\o(0) \over Z(T)Z(0)}\right)^{\textstyle{1\over 4}} \exp\biggl(
{-1 \over 2\h\sinh{\Th}(T)}
\sqrt{\o(T)\o(0) \over Z(T)Z(0)}\cr
&\times \Big\{\Big(
\sqrt{Z(0)\o(T) \over Z(T)\o(0)}q^{2}
+\sqrt{Z(T)\o(0)\over Z(0)\o(T)}q'^{2}\Big)\cosh{\Th}(T)
-2qq'^{2}\Big\}\cr
&-{i \over 2\h}\Big({Y(T) \over Z(T)}q^{2}-{Y(0) \over
Z(0)}q'^{2}\Big)\biggr)\quad . \cr}
}
\listrefs
\bye